\documentstyle[12pt,epsfig,color]{article}
\newcommand{\spacing}[1]{\renewcommand{\baselinestretch}{#1}\large\normalsize}
\spacing{1.5}
\begin{document}

\begin{center}
{\large\bf{Perspectives in Small Systems : Heavy Quarks and HBT Radii}} \\
\bigskip

{\small
Debasish Das$^{a}$\footnote{email : debasish.das@saha.ac.in,dev.deba@gmail.com}, 
\medskip

$^a$Saha Institute of Nuclear Physics, HBNI, 1/AF, Bidhannagar, Kolkata 700064, India\\
}

\end{center}
\date{\today}
\begin{abstract}

Recent observations of QGP-like phenomena in small collision systems like p+p and p+A collisions have
questioned our understanding of the basic paradigms of high energy heavy-ion physics.
A brief discussion of these new aspects in small systems which in turn influence our
understanding of hard probes like heavy quarks has been provided. Furthermore, a closer observation 
of the two-particle Hanbury-Brown Twiss (HBT) interferometry measurements provide new insights 
in our understanding of medium-like phenomena in small systems. An outlook of future goals and measurements
is also made.

\end{abstract}

\section{\bf{Introduction}}

Quarks are basic constituents of Quantum Chromo-Dynamics (QCD), and they interact via the
exchange of gluons~\cite{Fukushima:2011jc,Wilczek:1999ym}. QCD is a well established theory of
strong interactions~\cite{Kisslinger:2014uda,Shuryak:2008eq,Shuryak:1980tp,Bass:1998vz,Berges:1998nf}. 
From early calculations till very recent times Quark-Gluon Plasma (QGP) was illustrated and explained as a 
system ruled by the quarks and gluons which interact relatively weakly with each other. But the recent observations  
and theoretical developments now point towards the creation of a strongly coupled plasma liquid~\cite{Shuryak:2004cy,Heinz:2008tv}.

The only scope to probe such high and dense nuclear matter like QGP, in our laboratories
is possible via relativistic heavy-ion collisions~\cite{Adams:2005dq,Adcox:2004mh,Back:2004je,Arsene:2004fa,Abelev:2014ffa}.
What we observe in the laboratory are the particles like hadrons and leptons and
there are a large number of such individual probes~\cite{Armesto:2015ioy}. Among the key evidences for
the formation of a hot Quark-Gluon Plasma (QGP) in nucleus-nucleus (A+A) collisions~\cite{Busza:2018rrf}
at high collision energies is heavy-quark-antiquark pair (quarkonium)
suppression~\cite{Matsui:1986dk,Satz:1983jp,Andronic:2015wma} and bulk
collective effects~\cite{Voloshin:2008dg,Snellings:2011sz,Foka:2016vta},
together with their absence in proton-nucleus (p+A) control experiments~\cite{Fujii:2013gxa,Song:2017wtw,Nagle:2018nvi}.

Apart from understanding the measurements in nucleus-nucleus (A+A) collisions,
the study of the QGP requires reference measurements from systems in which the QGP is not
expected to be formed, such as p+p and p+A
collisions~\cite{Abelev:2014ffa,Abelev:2012qh,Ni:2018spv,MoreiraDeGodoy:2017wks}.
The transverse size of the overlap region is comparable to that of a single proton, in systems like p+p and p+A and
hence the formation of a hot and dense fluid-like medium or QGP is not
expected. The control measurements are however needed to characterize the extent
to which initial-state effects can be demarcated from the effects due to the
final-state interactions in the QGP. The p+A collisions due to the absence of a
produced medium are expected to isolate the nuclear effects from the initial
state of the hard-scattering process~\cite{Adams:2005dq,Adcox:2004mh,Back:2004je,Arsene:2004fa,Abelev:2014ffa,Acharya:2018yud}.
Also the system size evolution at freeze-out~\cite{Adare:2014vri}
can be studied from the small systems (like p+p and p+A), along
with the re-scattering effects~\cite{Kapusta:2005pt}.
Some of these assumptions are getting tested and understood carefully both in
Large Hadron Collider (LHC) and in Relativistic heavy-ion Collider (RHIC) data and
more can be possible in future at Electron Ion Collider (EIC).

The paper is organised to start with a brief introduction of small systems in
general and in section 2 we have a brief survey of existing results on
heavy quarks and an overview of the various issues that make a final 
conclusion on the nature of small systems very challenging.
Hanbury-Brown Twiss (HBT) interferometry~\cite{Lisa:2005dd} helps
us to decipher the system size and its evolution. This methodology, uses the Bose-Einstein quantum statistics 
to calculate the sizes of the particle emitting region. These sizes,  named as ``HBT radii ", are directly 
proportional to the width of the enhancement of two-particle correlation function at low relative momentum. 
Section 3 is devoted to the discussion
of why HBT is also an important tool towards the understanding of small systems.
Finally, we summarise by pointing out the challenges and future targets for such studies
with the small systems.


\section{\bf{Heavy Quarks}}
\label{sec:HQ_sec}

While traversing the QGP, heavy quarks are expected to lose a
significant fraction of their energy, which interact via the strong interaction~\cite{Baier:2000mf}.
Such energy loss may depend on the parton flavor. Those which interact via electroweak interactions do not experience such
energy loss~\cite{Horowitz:2008ig,Gossiaux:2010yx}. Quarkonia production in A+A and p+A,
as compared to p+p collisions has long been considered as a
celebrated tool to understand such production mechanisms. One such formalism is the study using the
nuclear modification factor~\cite{Ni:2018spv,MoreiraDeGodoy:2017wks,Zakharov:2012fp}.
If we consider the collisions between two nuclei, A and B, the nuclear modification
factor, $R_{AB}$ , is defined as the ratio of particle yield in AB collisions to those in p+p collisions
scaled by the average number of binary nucleon-nucleon collisions, $\langle N_{coll} \rangle$,
in AB collisions. It is given by,

\begin{equation}
  \label{eq_1}
R_{AB}(p_{T}) =
  \frac{d^{2}N_{AB}/dp_{T}dy_{\it c.m}}{\langle N_{coll} \rangle d^{2}N_{pp}/d{p_{T}}dy_{\it c.m}}
  =
\frac{d^{2}N_{AB}/dp_{T} dy_{\it c.m}}{\langle T_{AB} \rangle d^{2}\sigma_{pp}/d{p_{T}}dy_{\it c.m}}
\end{equation}

where $y_{\it c.m}$ is the rapidity calculated in the center-of-mass frame of the colliding nucleons, and
$\langle T_{AB} \rangle$, which is the nuclear overlap function, accounts for the nuclear collision geometry.
It is calculated using the Glauber model~\cite{Miller:2007ri}. If the nuclear collisions behave as
incoherent superpositions of nucleon-nucleon collisions, $R_{AB}$ is expected to be unity.
Also one expects $R_{AB}$ = 1 in the absence of initial and final state effects
in nuclear collisions. The nuclear modification factor, $R_{AB}$, equal to unity also means
no modification in heavy-ion collisions with respect to p+p collisions.

The bound states of heavy-quark-antiquark pairs (quarkonia) and heavy quarks
which are produced via initial hard scattering processes are crucial
probes for QGP~\cite{Kluberg:2005yh,Brambilla:2004wf,Das:2011bj,Mocsy:2013syh,Song:2015bja}.
Heavy quarks can access the initial anisotropy~\cite{Ollitrault:1992bk,Teaney:2010vd,Gardim:2011xv}
of the medium produced through high energy nucleus-nucleus collisions which can be realised in the final
measurement of quarkonium yield~\cite{Das:2018xel,Chen:2016mhl} in the experiments~\cite{Adamczyk:2012pw,Silvestre:2008tw,ALICE:2013xna,Khachatryan:2016ypw,Acharya:2017tgv,Acharya:2019hlv,CMS:2019uhg}. The modification
of the yield of the different quarkonium states is sensitive to
the temperature of the QGP~\cite{Mocsy:2013syh,Lansberg:2019adr,Kisslinger:2016ehx}. However, the suppression due to QGP must be
disentangled from Cold Nuclear Matter (CNM) effects (such as
nuclear modification of the parton distribution functions or break-up of the quarkonium state in CNM) 
which, as of now, are not exactly understood even at RHIC energies~\cite{Adamczyk:2013poh}.
The effects related to the presence of CNM can also modify the production of 
quarkonia in A+A collisions. CNM effects can be examined
in proton-nucleus (p+A) collisions, where the QGP is not expected to be formed~\cite{Vogt:2010aa}.

The p+p collisions are considered as a kind of elementary ones
or the reference process for detecting deconfined phase formation in A+A collisions.
It is believed that data for p+p and p+A collisions can help us to unravel
the details of the initial state effects in nuclear collisions.
The nuclear modification factor $R_{pA}$~\cite{Adamczyk:2013poh} is defined as ratio of p+A to p+p
cross-sections normalized to the average number of binary nucleon collisions.
The measurements by ALICE experiment~\cite{Adam:2015jsa} for the most peripheral p+Pb collisions
reveal almost no modification with respect to p+p collisions, within
the measured uncertainties. However, we see that the results
in central p+Pb collisions show sizeable nuclear effects.
Therefore, in order to interpret the heavy quark suppression in terms
of the nuclear modification factor, we must envisage to have a precise and firm
understanding of the p+A collisions~\cite{Vogt:2010aa,Arleo:2014oha}.

Are small systems like p+p collisions having flow effects? In recent studies by CMS~\cite{Khachatryan:2016txc}
we observe that within the experimental uncertainties, the multi-particle
cumulant~\cite{Bravina:2013ora} $v_{2}\{4\}$ and $v_{2}\{6\}$ values
in the high-multiplicity p+p collisions are consistent with each other and even
strikingly similar to what was observed previously in p+Pb and Pb+Pb collisions~\cite{Khachatryan:2015waa}.
The studies with charged particles (comprising of light quarks) show
strong evidence for the collective nature of the long-range correlations observed in p+p collisions
at $\sqrt{s}$ = 13 TeV at LHC~\cite{Khachatryan:2016txc}. Also at $\sqrt{s}$ = 13 TeV,
ATLAS experiment~\cite{Aad:2019aol} has measured elliptic flow ($v_{2}$) coefficients
for heavy-flavor decay muons in p+p collisions, along with
a separation between charm and bottom contributions. The bottom-decay muons show
$v_{2}$ values consistent with zero within statistical and systematic uncertainties,
while the charm-decay muons have significant non-zero $v_{2}$ values.
These results infer that bottom quarks have less elliptic flow ($v_{2}$) in high-multiplicity p+p collisions unlike the light and charm quarks.

For heavy-ions we see that the $v_{2}$ coefficient for bottomonia measurements are
compatible with zero as reported both by ALICE~\cite{Acharya:2019hlv} and CMS~\cite{CMS:2019uhg} experiments.
However, the smaller systems like p+A and p+p, have less deeply bound bottomonia states and hence
a comparatively larger chance to escape. This means that more states become measurable,
which is a positive feature. On the other hand, it also means that the escape mechanism
which underlies the anisotropic flow of bottomonia may become largely ineffective, in
particular for the $\Upsilon(1S)$. Accordingly, the measurement of a sizable
flow for $\Upsilon(1S)$ in small systems would probably hint at the
importance of initial-state correlations~\cite{Das:2018xel}.

We have established our formulation of nuclear modification factor (Eq.\ref{eq_1}) based on the normalization with such small systems,
considering them to be elementary and simple. Hence, the p+p collision system which have much smaller
overlap region needs to be monitored closely . It becomes clear that these recent experimental observations
in p+p collisions (like ATLAS for example) have challenged such views and we need to explore better formulations.
Alternately, a closer look into the difference of central and peripheral collisions
can be obtained by calculating the central-to-peripheral modification factor,

\begin{equation}
  \label{eq_2}
R_{CP} = \langle N^{peripheral}_{coll} \rangle  Yield^{central}_{Pb-Pb} / \langle N^{central}_{coll} \rangle Yield^{peripheral}_{Pb−Pb}.
\end{equation}

The ATLAS measurements~\cite{Aad:2010aa} for charm quarks in Pb+Pb (or A+A) collisions at
LHC energies show a significant decrease of $R_{CP}$ as a function of centrality.
For $R_{CP}$ the yields in the different centrality classes are divided by the corresponding
average number of binary collisions. As reference, the most peripheral bin is used for $R_{CP}$
considering that QGP formation is not expected in such peripheral A+A collisions.
If collective expansion occurs in peripheral A+A collisions, whether such formulations, like $R_{CP}$, will still 
hold needs to be understood. In peripheral A+A collisions, such collective aspects, can be examined by studying the system size 
evolution using two-particle momentum correlations (also known as HBT interferometry). 
The signature correlations between momentum and coordinate space result in collective flow. Understanding and measuring 
the system size via HBT interferometry, provides the most direct experimental access to such patterns. 
Thus a  convincing case for a bulk system in heavy-ion collisions depends crucially on such HBT measurements.


\section{\bf{HBT Radii}}
\label{sec:HBT_sec}

For understanding the bulk properties of QGP, the two-particle Hanbury-Brown-Twiss (HBT)
interferometry is an prime method towards the understanding of the
space-time structure of the particle emission sources produced in the
high energy heavy-ion collisions~\cite{Lisa:2005dd,Heinz:1996rw}.
The two-particle correlation function is defined as the ratio,

\begin{equation}
  \label{eq_3}
  C({\bf q})=\frac{A({\bf q})}{B({\bf q})}
  \end{equation}

 where A({\bf q}) is the measured distribution of pair momentum difference ${\bf q = p_{2} - p_{1}}$, and
B({\bf q}) is a similar distribution formed by using pairs of particles
from different events (event mixing). Since one can only deduce information about the 
relative separation of two particles with a given velocity, one actually estimates a subset of the entire source size, 
even if the correlation measurements cover all of the momentum space.  Thus, two-particle HBT correlations 
do not measure the size of the entire source. Instead, they measure the dimensions of the “region of homogeneity,” 
which in other words describe the size and shape of the phase space cloud of outgoing particles whose velocities have a specific magnitude and direction. 
If the collective expansion of the produced matter is strong, as we find in the case for central collisions, then the region of homogeneity 
is very much smaller than the entire source volume~\cite{Lisa:2005dd,Heinz:1996rw}.

The homogeneity length scales are calculated experimentally.  Following the often used experimental analysis 
techniques of recent times, we decompose the relative momentum of the pairs (shown in Eq.\ref{eq_3}) into the three projections, 
as ``out', ``side' and ``long", according to the Bertsch-Pratt convention~\cite{Lisa:2005dd}. These experimental measurements assume that the homogeneity regions
can be estimated via a Gaussian-profile ellipsoid in configuration space~\cite{Lisa:2005dd,Heinz:1996rw,Frodermann:2006sp,Ermakov:2017knq}.
Such studies result in a Gaussian two-particle momentum correlation function where a semi-analytic Gaussian fit to the relative momentum
dependence of the measured correlation function (Eq.\ref{eq_3}) is carried out.   The correlation functions were estimated in longitudinal 
co-moving system (LCMS), where $p_{z1} + p_{z2} = 0$. Assuming a Gaussian emission profile of the 
source, the correlations functions are fitted with the form as follows:

  \begin{equation}
   \label{eq_4}
C( {\bf q}, {\bf k_{T}} ) = 1 + \lambda( {\bf k_{T} }) \exp   [ - \sum_{i,j=out,side,long}  q_{i}q_{j} R^{2}_{ij} ( {\bf k_{T }})]
\end{equation}

where $\lambda$ denotes the fraction of correlated pairs and $q_{i}$ is the relative momentum of the pair in the i-direction. 
The longitudinal direction is along the beam-axis which corresponds to the long term, the outward direction is pointing along the transverse 
component of the average momentum $\bf k$ of a pair (${\bf k_{T} }= | {\bf p_{1T}} + {\bf p_{2T} }| / 2$) and the sideward direction is perpendicular to the other two i.e out and long. 
When experimentally the measurements are done using the pseudo-rapidity range of  $|\eta| < 1 $ , the effect of cross terms with $i \ne j$ in the HBT radii becomes negligible. 
Finally the system size parameters extracted from the Gaussian fits to the correlation function are named as ``HBT radii''~\cite{Lisa:2005dd,Heinz:1996rw}.
The HBT radii are related to regions of homogeneity where $R_{side}$ contains information about geometry, $R_{out}$ convolutes the information 
about geometry and emission duration and $R_{long}$ contains information about system lifetime. The relative momentum dependence of the HBT 
radii shows the dynamics of the collision system and enables to probe the different regions of the homogeneity~\cite{Lisa:2005dd,Heinz:1996rw,Frodermann:2006sp,Ermakov:2017knq}.

At freeze-out, the Hanbury-Brown-Twiss (HBT) interferometry measurements
provides the understanding of the source size and lifetime~\cite{Lisa:2005dd,Heinz:1996rw,Frodermann:2006sp}.
Such measurements in the elementary particle collisions infer about the
space-time distribution of the production points of the hadrons. For the small
systems the re-scattering effects are usually negligible as the number of
particles produced are small and also the source size~\cite{Florkowski:2016xbk}.
The source here then may be considered to be transparent. It is often understood that
HBT measures the distribution of last scatterings~\cite{Wong:2004gm}. For heavy-ion
collisions re-scattering effects after production becomes significant.
The source size is significantly large, both due to the number of
nucleons which is large and also because the number of various
produced particles, which is large. For A+A collisions the source
is neither opaque nor transparent~\cite{Kapusta:2005pt,Wong:2004gm}.

The HBT measurements showing the dependence on transverse mass
and collision centrality in A+A systems, are known to exhibit features
that point to the influence of the final-state re-scattering effects~\cite{Kapusta:2005pt}.
Also the momentum dependence of the femtoscopic scales reveal the
coordinate-space aspects of collective expansion or flow~\cite{Lisa:2005dd,Heinz:1996rw,Voloshin:1995mc}.
Comparing the HBT measurements in A+A and small systems (like p+A and p+p)
can help in understanding the impact that the final-state re-scattering
effects has in the small systems. Similar measurements for d+A and p+A
collisions (or small systems) will infer an important avenue to
independently examine the role of final-state interactions in the reaction dynamics
for these systems. An observed similarity between the characteristic
patterns for the space–time extent of A+A and p+A (or d+A) collisions including the p+p collisions
at higher energies like LHC, would give a strong indication for
the influence of final-state re-scattering effects in the small system collisions~\cite{Adare:2014vri,Aamodt:2011kd}.

The experimental measurements performed for heavy-ion collisions at RHIC energies
have revealed two very important aspects. Firstly, the three-dimensional HBT radii
(also known as $R_{out}$, $R_{side}$ and $R_{long}$) in the Longitudinally Co-Moving System (LCMS)~\cite{Lisa:2005dd} are
observed to scale linearly with the cube root of charged particle
pseudo-rapidity density $\langle dN_{ch}/d\eta \rangle^{1/3}$~\cite{Das:2007ny}. At top energies in RHIC
the HBT radii of small systems (like p+p and d+A) scale linearly with $\langle dN_{ch}/d\eta \rangle^{1/3}$
as also seen for heavy-ion systems like Au+Au and Cu+Cu~\cite{Das:2007ny}. 
At higher LHC energies for the systems like p+p and p+A, we observe for ALICE measurements~\cite{Adam:2015pya}, that the HBT radii scale
roughly with $\langle dN_{ch}/d\eta \rangle^{1/3}$ not only with A+A collisions~\cite{Adam:2015vna} at LHC,
but also across with other A+A collision energies and systems of RHIC and Super Proton Synchrotron (SPS). 
Such universal scaling pattern across several collision energies and systems show
hydrodynamic-like collective expansion driven by final-state re-scattering effects~\cite{Kapusta:2005pt,Wong:2004gm}.

But at LHC with larger data-sets do we see something special in small systems?
With copious p+p datasets at LHC energies, we now find a prominent different scaling behavior
when compared with A+A results, although both exhibit linearity with $\langle dN_{ch}/d\eta \rangle^{1/3}$.
At LHC energies the p+Pb (or p+A) radii are comparable with p+p collisions at low multiplicities.
But with increasing multiplicity at LHC regime, the radii for the two systems start to diverge, which is a new addition to HBT measurements. 
However, the results from one-dimensional averaged radii in p+p, p+A (or p+Pb) and A+A (or Pb+Pb) collisions using the 3-pion cumulant
correlations technique by ALICE~\cite{Abelev:2014pja} also strengthens the universal observations that the multiplicity scaling for
p+A lies between p+p and A+A trends. The smaller radii in p+A as compared to A+A collisions show the importance of different 
initial conditions on the final-state. It can further indicate sizeable collective expansion already in peripheral A+A collisions.

The second very important aspect of the HBT radii is the observed decrease
with increasing pair transverse momentum. Such behavior indicates
the presence of the strong collective radial flow in the medium.
Collective expansion results in position-momentum correlations in both
transverse and longitudinal directions~\cite{Lisa:2005dd,Heinz:1996rw,Voloshin:1995mc}.

The smaller system like Cu+Cu overlap with peripheral Au+Au. Hence to inspect such flow effects in peripheral
Au+Au (or A+A) collisions, the pion pair momentum dependences
of the HBT radii were measured by STAR experiment at RHIC~\cite{Abelev:2009tp}. Although the
individual HBT radii decrease significantly with increasing pair
transverse momentum, the radii ratios ( $Au+Au/Cu+Cu$ ) show that the HBT radii
for Au+Au and Cu+Cu collisions share a common dependence, in terms of
pair transverse momenta. As reported by ALICE~\cite{Adam:2015pya} and CMS~\cite{Sirunyan:2017ies}
experiments at LHC energies, such decrease of radii with pair transverse momenta is observed both in
p+p and p+Pb collisions (or small systems), like what has been observed for the heavy-ions (or A+A collisions).

\section{Summary and Outlook}

Experimental results in small colliding systems showing emanation of QGP-like phenomena
has generated avid interest in recent times, including critical discussions based on their
measurements on heavy quark suppression and collective behavior. 
In high-energy A+A collisions, bottomonium suppression~\cite{Adare:2014hje,Chatrchyan:2011pe,Abelev:2014nua,Chatrchyan:2012lxa,Das:2008zzg}
can be attributed as a ``golden probe'' for QGP creation, since CNM effects are expected to be small in Upsilon ($\Upsilon$) production~\cite{Das:2018xel}.
Due to the larger mass of the bottomonium states compared to the charmonium ones, the measurement of bottomonia production in 
proton-nucleus collisions~\cite{Abelev:2014oea,Aaij:2014mza,Aaij:2018scz} allows a study of CNM effects in a different kinematic regime, 
and therefore complementing the J/$\Psi$ studies~\cite{Adam:2015jsa,Abelev:2013yxa,Aaij:2013zxa,Aaij:2017cqq}.
Determining the meson propagation is an important aspect towards the understanding of
quarkonium suppression in heavy-ion collisions. There are several such possible signatures of the 
slow velocity of quarkonium mesons. Also there are propositions that quarkonium HBT interferometry~\cite{Ejaz:2007hg} may in principle
be able to find signatures of a depressed meson velocity. However, while performing the calculation of energy loss in small systems a theoretical 
problem arises, as all of the standard pQCD energy loss models make the explicit assumption that the QGP system is large~\cite{Hambrock:2018sim}.
Recent research on more heavier quarks like top~\cite{Apolinario:2017sob,Kisslinger:2019vvc}, which has a short lifetime,
propose its utility to probe the temporal structure of the QGP evolution.
Since the top quarks decay mostly within the strongly interacting medium, the latest
experimental observations~\cite{Sirunyan:2017xku,CMS:2019fmm} may further establish them, as a new probe
to study the mechanisms of medium-induced parton energy loss. How much they can
help in solving the anomalies in small systems, is what we need to explore further
in forthcoming LHC regime of Run-3 and 4.

Exceptionally high-multiplicity p+p collisions are expected in Run-3 and 4 at LHC.
Such higher multiplicities will help us to bridge the gap between
the p+p and heavy-ion collisions, with better detector upgrades in LHC experiments~\cite{Citron:2018lsq}.
Better picture will be also available from p+Pb collisions as seen with recent results on 
multiplicity dependent quarkonium measurements from ALICE~\cite{Acharya:2020giw}.
However, further investigations and the increasing precision in LHC Run-3 and 4 will be necessary to improve the 
understanding of the interplay between hard and soft processes in high multiplicity p+p collisions.
The rare probes can be explored in further detail in small systems with much larger data-sets.

The proposed EIC~\cite{Accardi:2012qut}, with a lepton beam on a variety of light and heavy nuclei at a range of center-of-mass energies,
will offer a more cleaner experimental scenario (no pile-up, full kinematic reconstruction) and hence enable precision 
measurements of the flavor-separated partonic structure of nuclei, comparable to similar measurements performed on protons. 
Quarkonium production in eA collisions where one can understand the interactions with nuclear matter along with the formation of 
quarkonia in a nuclear medium are among the few main objectives of EIC. The study of nuclear effects in quarkonium production is a relatively 
new area of nuclear physics, where EIC measurements will be important as well as significant in taking the subject forward. 
The comparison between the two-particle elliptic flow in p+p and p+A collisions and further observables in e+A 
collisions~\cite{Hagiwara:2017ofm} will be paramount towards the understanding of the gluon dynamics under extreme conditions.


\bibliography{apssamp}

\begin{thebibliography}{}

\bibitem{Fukushima:2011jc} 
  K.~Fukushima,
  J.\ Phys.\ G {\bf 39}, 013101 (2012)
  doi:10.1088/0954-3899/39/1/013101
  [arXiv:1108.2939 [hep-ph]].


\bibitem{Wilczek:1999ym}
  F.~Wilczek,
  arXiv:hep-ph/0003183.

\bibitem{Kisslinger:2014uda} 
  L.~S.~Kisslinger and D.~Das,
  Int.\ J.\ Mod.\ Phys.\ A {\bf 31}, no. 07, 1630010 (2016)
  doi:10.1142/S0217751X16300106
  [arXiv:1411.3680 [hep-ph]].


\bibitem{Shuryak:2008eq} 
  E.~Shuryak,
  Prog.\ Part.\ Nucl.\ Phys.\  {\bf 62}, 48 (2009)
  doi:10.1016/j.ppnp.2008.09.001
  [arXiv:0807.3033 [hep-ph]].

\bibitem{Shuryak:1980tp}
  E.~V.~Shuryak,
  Phys.\ Rept.\  {\bf 61}, 71 (1980).

\bibitem{Bass:1998vz} 
  S.~A.~Bass, M.~Gyulassy, H.~Stoecker and W.~Greiner,
  J.\ Phys.\ G {\bf 25}, R1 (1999)
  doi:10.1088/0954-3899/25/3/013
  [hep-ph/9810281].
  

  
\bibitem{Berges:1998nf}
  J.~Berges,
  arXiv:hep-ph/9902419.

  
 
\bibitem{Shuryak:2004cy}
E.~V.~Shuryak,
Nucl. Phys. A \textbf{750}, 64-83 (2005)
doi:10.1016/j.nuclphysa.2004.10.022
[arXiv:hep-ph/0405066 [hep-ph]].

 
\bibitem{Heinz:2008tv}
U.~W.~Heinz,
J. Phys. A \textbf{42}, 214003 (2009)
doi:10.1088/1751-8113/42/21/214003
[arXiv:0810.5529 [nucl-th]].


\bibitem{Adams:2005dq} 
  J.~Adams {\it et al.} [STAR Collaboration],
  Nucl.\ Phys.\ A {\bf 757}, 102 (2005)
  doi:10.1016/j.nuclphysa.2005.03.085
  [nucl-ex/0501009].

\bibitem{Adcox:2004mh} 
  K.~Adcox {\it et al.} [PHENIX Collaboration],
  Nucl.\ Phys.\ A {\bf 757}, 184 (2005)
  doi:10.1016/j.nuclphysa.2005.03.086
  [nucl-ex/0410003].


\bibitem{Back:2004je} 
  B.~B.~Back {\it et al.},
  Nucl.\ Phys.\ A {\bf 757}, 28 (2005)
  doi:10.1016/j.nuclphysa.2005.03.084
  [nucl-ex/0410022].


\bibitem{Arsene:2004fa} 
  I.~Arsene {\it et al.} [BRAHMS Collaboration],
  Nucl.\ Phys.\ A {\bf 757}, 1 (2005)
  doi:10.1016/j.nuclphysa.2005.02.130
  [nucl-ex/0410020].


\bibitem{Abelev:2014ffa} 
  B.~B.~Abelev {\it et al.} [ALICE Collaboration],
  Int.\ J.\ Mod.\ Phys.\ A {\bf 29}, 1430044 (2014)
  doi:10.1142/S0217751X14300440
  [arXiv:1402.4476 [nucl-ex]].

\bibitem{Armesto:2015ioy} 
  N.~Armesto and E.~Scomparin,
  Eur.\ Phys.\ J.\ Plus {\bf 131}, no. 3, 52 (2016)
  doi:10.1140/epjp/i2016-16052-4
  [arXiv:1511.02151 [nucl-ex]].

\bibitem{Busza:2018rrf} 
  W.~Busza, K.~Rajagopal and W.~van der Schee,
  Ann.\ Rev.\ Nucl.\ Part.\ Sci.\  {\bf 68}, 339 (2018)
  doi:10.1146/annurev-nucl-101917-020852
  [arXiv:1802.04801 [hep-ph]].



\bibitem{Matsui:1986dk}
  T.~Matsui and H.~Satz,
 Phys.\ Lett.\ B \textbf{178} (1986) 416.



  
\bibitem{Satz:1983jp}
  H.~Satz,
  Nucl.\ Phys.\ A {\bf 418}, 447C (1984).



\bibitem{Andronic:2015wma} 
  A.~Andronic {\it et al.},
  Eur.\ Phys.\ J.\ C {\bf 76}, no. 3, 107 (2016)
  doi:10.1140/epjc/s10052-015-3819-5
  [arXiv:1506.03981 [nucl-ex]].



\bibitem{Voloshin:2008dg} 
  S.~A.~Voloshin, A.~M.~Poskanzer and R.~Snellings,
  arXiv:0809.2949 [nucl-ex].



\bibitem{Snellings:2011sz} 
  R.~Snellings,
  New J.\ Phys.\  {\bf 13}, 055008 (2011)
  doi:10.1088/1367-2630/13/5/055008
  [arXiv:1102.3010 [nucl-ex]].


\bibitem{Foka:2016vta} 
  P.~Foka and M.~A.~Janik,
  Rev.\ Phys.\  {\bf 1}, 154 (2016)
  doi:10.1016/j.revip.2016.11.002
  [arXiv:1702.07233 [hep-ex]].

\bibitem{Fujii:2013gxa} 
  H.~Fujii and K.~Watanabe,\newline
  Nucl.\ Phys.\ A {\bf 915}, 1 (2013)
  doi:10.1016/j.nuclphysa.2013.06.011
  [arXiv:1304.2221 [hep-ph]].


\bibitem{Song:2017wtw} 
  H.~Song, Y.~Zhou and K.~Gajdosova,
  Nucl.\ Sci.\ Tech.\  {\bf 28}, no. 7, 99 (2017)
  doi:10.1007/s41365-017-0245-4
  [arXiv:1703.00670 [nucl-th]].

\bibitem{Nagle:2018nvi} 
  J.~L.~Nagle and W.~A.~Zajc,
  Ann.\ Rev.\ Nucl.\ Part.\ Sci.\  {\bf 68}, 211 (2018)
  doi:10.1146/annurev-nucl-101916-123209
  [arXiv:1801.03477 [nucl-ex]].



\bibitem{Abelev:2012qh} 
  B.~Abelev {\it et al.} [ALICE Collaboration],
  Phys.\ Rev.\ Lett.\  {\bf 109}, 112301 (2012)
  doi:10.1103/PhysRevLett.109.112301
  [arXiv:1205.6443 [hep-ex]].


\bibitem{Ni:2018spv} 
  H.~Ni [CMS Collaboration],
  J.\ Phys.\ Conf.\ Ser.\  {\bf 1070}, no. 1, 012009 (2018).
  doi:10.1088/1742-6596/1070/1/012009

\bibitem{MoreiraDeGodoy:2017wks} 
  D.~Moreira De Godoy [ALICE Collaboration],
  Nucl.\ Phys.\ A {\bf 967}, 636 (2017)
  doi:10.1016/j.nuclphysa.2017.05.050
  [arXiv:1705.02800 [nucl-ex]].


\bibitem{Acharya:2018yud} 
  S.~Acharya {\it et al.} [ALICE Collaboration],
  Eur.\ Phys.\ J.\ C {\bf 78}, no. 6, 466 (2018)
  doi:10.1140/epjc/s10052-018-5881-2
  [arXiv:1802.00765 [nucl-ex]].

  
\bibitem{Adare:2014vri} 
  N.~N.~Ajitanand {\it et al.} [PHENIX Collaboration],
  Nucl.\ Phys.\ A {\bf 931}, 1082 (2014)
  doi:10.1016/j.nuclphysa.2014.08.054
  [arXiv:1404.5291 [nucl-ex]].


\bibitem{Kapusta:2005pt} 
  J.~I.~Kapusta and Y.~Li, \newline
  Phys.\ Rev.\ C {\bf 72}, 064902 (2005)
  doi:10.1103/PhysRevC.72.064902
  [nucl-th/0503075].




\bibitem{Lisa:2005dd} 
  M.~A.~Lisa, S.~Pratt, R.~Soltz and U.~Wiedemann,
  Ann.\ Rev.\ Nucl.\ Part.\ Sci.\  {\bf 55}, 357 (2005)
  doi:10.1146/annurev.nucl.55.090704.151533
  [nucl-ex/0505014].



\bibitem{Baier:2000mf} 
  R.~Baier, D.~Schiff and B.~G.~Zakharov,
  Ann.\ Rev.\ Nucl.\ Part.\ Sci.\  {\bf 50}, 37 (2000)
  doi:10.1146/annurev.nucl.50.1.37
  [hep-ph/0002198].

\bibitem{Horowitz:2008ig} 
  W.~A.~Horowitz and M.~Gyulassy,
  J.\ Phys.\ G {\bf 35}, 104152 (2008)
  doi:10.1088/0954-3899/35/10/104152
  [arXiv:0804.4330 [hep-ph]].


\bibitem{Gossiaux:2010yx} 
  P.~B.~Gossiaux, J.~Aichelin, T.~Gousset and V.~Guiho,
  J.\ Phys.\ G {\bf 37}, 094019 (2010)
  doi:10.1088/0954-3899/37/9/094019
  [arXiv:1001.4166 [hep-ph]].

  

\bibitem{Zakharov:2012fp} 
  B.~G.~Zakharov,
  JETP Lett.\  {\bf 96}, 616 (2013)
  doi:10.1134/S002136401222016X
  [arXiv:1210.4148 [hep-ph]].


\bibitem{Miller:2007ri} 
  M.~L.~Miller, K.~Reygers, S.~J.~Sanders and P.~Steinberg,
  Ann.\ Rev.\ Nucl.\ Part.\ Sci.\  {\bf 57}, 205 (2007)
  doi:10.1146/annurev.nucl.57.090506.123020
  [nucl-ex/0701025].


\bibitem{Kluberg:2005yh}
  L.~Kluberg,
  Eur.\ Phys.\ J.\  C {\bf 43} (2005) 145.
\bibitem{Brambilla:2004wf}
  N.~Brambilla {\it et al.},
CERN Yellow Rep. 2005-005, [{\sf arXiv:hep-ph/0412158}].

\bibitem{Das:2011bj} 
  D.~Das [ALICE Collaboration],\\
  Nucl.\ Phys.\ A {\bf 862-863}, 223 (2011)
  doi:10.1016/j.nuclphysa.2011.05.044
  [arXiv:1102.2071 [nucl-ex]].



\bibitem{Mocsy:2013syh} 
  A.~Mocsy, P.~Petreczky and M.~Strickland,
  Int.\ J.\ Mod.\ Phys.\ A {\bf 28}, 1340012 (2013)
  doi:10.1142/S0217751X13400125
  [arXiv:1302.2180 [hep-ph]].




\bibitem{Song:2015bja} 
  T.~Song, C.~M.~Ko and S.~H.~Lee,
  Phys.\ Rev.\ C {\bf 91}, no. 4, 044909 (2015)
  doi:10.1103/PhysRevC.91.044909
  [arXiv:1502.05734 [nucl-th]].



\bibitem{Ollitrault:1992bk}
  J.~Y.~Ollitrault,
  Phys.\ Rev.\ D {\bf 46} (1992) 229.


\bibitem{Teaney:2010vd}
  D.~Teaney, L.~Yan,
  Phys.\ Rev.\  {\bf C83}, 064904 (2011).


\bibitem{Gardim:2011xv} 
  F.~G.~Gardim, F.~Grassi, M.~Luzum and J.~Y.~Ollitrault,
  Phys.\ Rev.\ C {\bf 85}, 024908 (2012)
  doi:10.1103/PhysRevC.85.024908
  [arXiv:1111.6538 [nucl-th]].


\bibitem{Das:2018xel} 
  D.~Das and N.~Dutta,
  Int.\ J.\ Mod.\ Phys.\ A {\bf 33}, no. 16, 1850092 (2018)
  doi:10.1142/S0217751X18500926
  [arXiv:1802.00414 [nucl-ex]].


\bibitem{Chen:2016mhl} 
  B.~Chen,
  Phys.Rev.C {\bf 95}, no. 3, 034908 (2017)
  doi:10.1103/PhysRevC.95.034908
  [arXiv:1608.02173 [nucl-th]].


\bibitem{Adamczyk:2012pw} 
  L.~Adamczyk {\it et al.} [STAR Collaboration],
  Phys.\ Rev.\ Lett.\  {\bf 111}, no. 5, 052301 (2013)
  doi:10.1103/PhysRevLett.111.052301
  [arXiv:1212.3304 [nucl-ex]].

\bibitem{Silvestre:2008tw} 
  C.~Silvestre [PHENIX Collaboration],
  J.\ Phys.\ G {\bf 35}, 104136 (2008)
  doi:10.1088/0954-3899/35/10/104136
  [arXiv:0806.0475 [nucl-ex]].


\bibitem{ALICE:2013xna} 
  E.~Abbas {\it et al.} [ALICE Collaboration],
  Phys.\ Rev.\ Lett.\  {\bf 111}, 162301 (2013)
  doi:10.1103/PhysRevLett.111.162301
  [arXiv:1303.5880 [nucl-ex]].


\bibitem{Khachatryan:2016ypw} 
  V.~Khachatryan {\it et al.} [CMS Collaboration],
  Eur.\ Phys.\ J.\ C {\bf 77}, no. 4, 252 (2017)
  doi:10.1140/epjc/s10052-017-4781-1
  [arXiv:1610.00613 [nucl-ex]].


\bibitem{Acharya:2017tgv} 
  S.~Acharya {\it et al.} [ALICE Collaboration],
  Phys.\ Rev.\ Lett.\  {\bf 119}, no. 24, 242301 (2017)
  doi:10.1103/PhysRevLett.119.242301
  [arXiv:1709.05260 [nucl-ex]].

  
\bibitem{Acharya:2019hlv} 
  S.~Acharya {\it et al.} [ALICE Collaboration],
  Phys.\ Rev.\ Lett.\  {\bf 123}, no. 19, 192301 (2019)
  doi:10.1103/PhysRevLett.123.192301
  [arXiv:1907.03169 [nucl-ex]].

\bibitem{CMS:2019uhg} 
  CMS Collaboration [CMS Collaboration],
  CMS-PAS-HIN-19-002.

  
\bibitem{Lansberg:2019adr} 
  J.~P.~Lansberg,
  arXiv:1903.09185 [hep-ph].



\bibitem{Kisslinger:2016ehx} 
  L.~S.~Kisslinger and D.~Das,
  JHEP {\bf 1709}, 105 (2017)
  doi:10.1007/JHEP09(2017)105
  [arXiv:1612.02269 [hep-ph]].


\bibitem{Adamczyk:2013poh} 
  L.~Adamczyk {\it et al.} [STAR Collaboration],
  Phys.\ Lett.\ B {\bf 735}, 127 (2014)
  Erratum: [Phys.\ Lett.\ B {\bf 743}, 537 (2015)]
  doi:10.1016/j.physletb.2014.06.028, 10.1016/j.physletb.2015.01.046
  [arXiv:1312.3675 [nucl-ex]].


\bibitem{Vogt:2010aa} 
  R.~Vogt,
  Phys.\ Rev.\ C {\bf 81}, 044903 (2010)
  doi:10.1103/PhysRevC.81.044903
  [arXiv:1003.3497 [hep-ph]].


\bibitem{Adam:2015jsa} 
  J.~Adam {\it et al.} [ALICE Collaboration],
  JHEP {\bf 1511}, 127 (2015)
  doi:10.1007/JHEP11(2015)127
  [arXiv:1506.08808 [nucl-ex]].


\bibitem{Arleo:2014oha} 
  F.~Arleo and S.~Peigné,
  JHEP {\bf 1410}, 073 (2014)
  doi:10.1007/JHEP10(2014)073
  [arXiv:1407.5054 [hep-ph]].

\bibitem{Khachatryan:2016txc} 
  V.~Khachatryan {\it et al.} [CMS Collaboration],
  Phys.\ Lett.\ B {\bf 765}, 193 (2017)
  doi:10.1016/j.physletb.2016.12.009
  [arXiv:1606.06198 [nucl-ex]].



\bibitem{Bravina:2013ora} 
  L.~V.~Bravina { \it et al.}, \\
  Phys.\ Rev.\ C {\bf 89}, no. 2, 024909 (2014)
  doi:10.1103/PhysRevC.89.024909
  [arXiv:1311.0747 [hep-ph]].


  
\bibitem{Khachatryan:2015waa} 
  V.~Khachatryan {\it et al.} [CMS Collaboration],
  Phys.\ Rev.\ Lett.\  {\bf 115}, no. 1, 012301 (2015)
  doi:10.1103/PhysRevLett.115.012301
  [arXiv:1502.05382 [nucl-ex]].



\bibitem{Aad:2019aol}
G.~Aad \textit{et al.} [ATLAS],
Phys. Rev. Lett. \textbf{124}, no.8, 082301 (2020)
doi:10.1103/PhysRevLett.124.082301
[arXiv:1909.01650 [nucl-ex]].
  
  

\bibitem{Aad:2010aa} 
  G.~Aad {\it et al.} [ATLAS Collaboration],
  Phys.\ Lett.\ B {\bf 697}, 294 (2011)
  doi:10.1016/j.physletb.2011.02.006
  [arXiv:1012.5419 [hep-ex]].

\bibitem{Heinz:1996rw} 
  U.~W.~Heinz,
  Nucl.\ Phys.\ A {\bf 610}, 264C (1996)
  doi:10.1016/S0375-9474(96)00361-2
  [nucl-th/9608002].

\bibitem{Frodermann:2006sp} 
  E.~Frodermann, U.~Heinz and M.~A.~Lisa,
  Phys.\ Rev.\ C {\bf 73}, 044908 (2006)
  doi:10.1103/PhysRevC.73.044908
  [nucl-th/0602023].


\bibitem{Ermakov:2017knq}
N.~Ermakov and G.~Nigmatkulov,
J. Phys. Conf. Ser. \textbf{798}, no.1, 012055 (2017)
doi:10.1088/1742-6596/798/1/012055
[arXiv:1806.03550 [nucl-ex]].

  
\bibitem{Florkowski:2016xbk} 
  W.~Florkowski,
  Acta Phys.\ Polon.\ B {\bf 47}, 2241 (2016)
  doi:10.5506/APhysPolB.47.2241
  [arXiv:1603.06418 [nucl-th]].

\bibitem{Wong:2004gm} 
  C.~Y.~Wong,
  J.\ Phys.\ G {\bf 30}, S1053 (2004)
  doi:10.1088/0954-3899/30/8/057
  [hep-ph/0403025].


\bibitem{Voloshin:1995mc} 
  S.~A.~Voloshin and W.~E.~Cleland,
  Phys.\ Rev.\ C {\bf 53}, 896 (1996)
  doi:10.1103/PhysRevC.53.896
  [nucl-th/9509025].

\bibitem{Aamodt:2011kd} 
  K.~Aamodt {\it et al.} [ALICE Collaboration],
  Phys.\ Rev.\ D {\bf 84}, 112004 (2011)
  doi:10.1103/PhysRevD.84.112004
  [arXiv:1101.3665 [hep-ex]].



\bibitem{Das:2007ny} 
  D.~Das[STAR Collaboration],
  Int.\ J.\ Mod.\ Phys.\ E {\bf 16}, 1883 (2007)
  doi:10.1142/S0218301307007179
  [nucl-ex/0702047 [NUCL-EX]].


\bibitem{Adam:2015pya} 
  J.~Adam {\it et al.} [ALICE Collaboration],
  Phys.\ Rev.\ C {\bf 91}, 034906 (2015)
  doi:10.1103/PhysRevC.91.034906
  [arXiv:1502.00559 [nucl-ex]].

\bibitem{Adam:2015vna} 
  J.~Adam {\it et al.} [ALICE Collaboration],
  Phys.\ Rev.\ C {\bf 93}, no. 2, 024905 (2016)
  doi:10.1103/PhysRevC.93.024905
  [arXiv:1507.06842 [nucl-ex]].


\bibitem{Abelev:2014pja} 
  B.~B.~Abelev {\it et al.} [ALICE Collaboration],
  Phys.\ Lett.\ B {\bf 739}, 139 (2014)
  doi:10.1016/j.physletb.2014.10.034
  [arXiv:1404.1194 [nucl-ex]].

\bibitem{Abelev:2009tp} 
  B.~I.~Abelev {\it et al.} [STAR Collaboration],
  Phys.\ Rev.\ C {\bf 80}, 024905 (2009)
  doi:10.1103/PhysRevC.80.024905
  [arXiv:0903.1296 [nucl-ex]].




\bibitem{Sirunyan:2017ies} 
  A.~M.~Sirunyan {\it et al.} [CMS Collaboration],
  Phys.\ Rev.\ C {\bf 97}, no. 6, 064912 (2018)
  doi:10.1103/PhysRevC.97.064912
  [arXiv:1712.07198 [hep-ex]].




  
\bibitem{Adare:2014hje} 
  A.~Adare {\it et al.} [PHENIX Collaboration],
  Phys.\ Rev.\ C {\bf 91}, no. 2, 024913 (2015)
  doi:10.1103/PhysRevC.91.024913
  [arXiv:1404.2246 [nucl-ex]].



\bibitem{Chatrchyan:2011pe} 
  S.~Chatrchyan {\it et al.} [CMS Collaboration],
  Phys.\ Rev.\ Lett.\  {\bf 107}, 052302 (2011)
  doi:10.1103/PhysRevLett.107.052302
  [arXiv:1105.4894 [nucl-ex]].




\bibitem{Abelev:2014nua} 
  B.~B.~Abelev {\it et al.} [ALICE Collaboration],
  Phys.\ Lett.\ B {\bf 738}, 361 (2014)
  doi:10.1016/j.physletb.2014.10.001
  [arXiv:1405.4493 [nucl-ex]].


\bibitem{Chatrchyan:2012lxa} 
  S.~Chatrchyan {\it et al.} [CMS Collaboration],
  Phys.\ Rev.\ Lett.\  {\bf 109}, 222301 (2012)
  doi:10.1103/PhysRevLett.109.222301
  [arXiv:1208.2826 [nucl-ex]].


  
\bibitem{Das:2008zzg} 
  D.~Das[STAR Collaboration],
  Eur.\ Phys.\ J.\ C {\bf 62}, 95 (2009).
  doi:10.1140/epjc/s10052-009-0937-y


\bibitem{Abelev:2014oea} 
  B.~B.~Abelev {\it et al.} [ALICE Collaboration],
  Phys.\ Lett.\ B {\bf 740}, 105 (2015)
  doi:10.1016/j.physletb.2014.11.041
  [arXiv:1410.2234 [nucl-ex]].

\bibitem{Aaij:2014mza} R.Aaij {\it et al.} [LHCb Collaboration],\newline
  JHEP {\bf 1407}, 094 (2014)
  doi:10.1007/JHEP07(2014)094
  [arXiv:1405.5152 [nucl-ex]].


\bibitem{Aaij:2018scz} R.Aaij {\it et al.} [LHCb Collaboration],\newline
  JHEP {\bf 1811}, 194 (2018)
  doi:10.1007/JHEP11(2018)194
  [arXiv:1810.07655 [hep-ex]].



\bibitem{Abelev:2013yxa}  
  B.~B.~Abelev {\it et al.} [ALICE Collaboration],\\
  JHEP {\bf 1402}, 073 (2014)
  doi:10.1007/JHEP02(2014)073
  [arXiv:1308.6726 [nucl-ex]].




\bibitem{Aaij:2013zxa} R.~Aaij {\it et al.} [LHCb Collaboration],\\
  JHEP {\bf 1402}, 072 (2014)
  doi:10.1007/JHEP02(2014)072
  [arXiv:1308.6729 [nucl-ex]].


\bibitem{Aaij:2017cqq} 
  R.~Aaij {\it et al.} [LHCb Collaboration],
  Phys.\ Lett.\ B {\bf 774}, 159 (2017)
  doi:10.1016/j.physletb.2017.09.058
  [arXiv:1706.07122 [hep-ex]].

  
\bibitem{Ejaz:2007hg} 
  Q.~J.~Ejaz, T.~Faulkner, H.~Liu, K.~Rajagopal and U.~A.~Wiedemann,
  JHEP {\bf 0804}, 089 (2008)
  doi:10.1088/1126-6708/2008/04/089
  [arXiv:0712.0590 [hep-th]].

\bibitem{Hambrock:2018sim} 
  R.~Hambrock and W.~A.~Horowitz,
  EPJ Web Conf.\  {\bf 171}, 18002 (2018)
  doi:10.1051/epjconf/201817118002
  [arXiv:1802.02442 [hep-ph]].

  

\bibitem{Apolinario:2017sob} 
  L.~Apolinário, J.~G.~Milhano, G.~P.~Salam and C.~A.~Salgado,
  Phys.\ Rev.\ Lett.\  {\bf 120}, no. 23, 232301 (2018)
  doi:10.1103/PhysRevLett.120.232301
  [arXiv:1711.03105 [hep-ph]].


\bibitem{Kisslinger:2019vvc} 
  L.~S.~Kisslinger and D.~Das,
  Mod.\ Phys.\ Lett.\ A {\bf 38}, 1950353 (2020)
  [Mod.\ Phys.\ Lett.\ A {\bf 34}, 1950353 (2019)]
  doi:10.1142/S021773231950353X
  [arXiv:1910.11101 [physics.gen-ph]].


\bibitem{Sirunyan:2017xku} 
  A.~M.~Sirunyan {\it et al.} [CMS Collaboration],
  Phys.\ Rev.\ Lett.\  {\bf 119}, no. 24, 242001 (2017)
  doi:10.1103/PhysRevLett.119.242001
  [arXiv:1709.07411 [nucl-ex]].

\bibitem{CMS:2019fmm} 
  CMS Collaboration [CMS Collaboration],
  CMS-PAS-HIN-19-001.



\bibitem{Citron:2018lsq} 
  Z.~Citron {\it et al.},
  CERN Yellow Rep.\ Monogr.\ , 1159 (2019)
  doi:10.23731/CYRM-2019-007.1159
  [arXiv:1812.06772 [hep-ph]].


\bibitem{Acharya:2020giw}
S.~Acharya \textit{et al.} [ALICE],
JHEP \textbf{09}, 162 (2020)
doi:10.1007/JHEP09(2020)162
[arXiv:2004.12673 [nucl-ex]].



\bibitem{Accardi:2012qut} 
  A.~Accardi {\it et al.},
  Eur.\ Phys.\ J.\ A {\bf 52}, no. 9, 268 (2016)
  doi:10.1140/epja/i2016-16268-9
  [arXiv:1212.1701 [nucl-ex]].

\bibitem{Hagiwara:2017ofm}
  Y.~Hagiwara, Y.~Hatta, B.~W.~Xiao and F.~Yuan,
  Phys.\ Lett.\ B {\bf 771}, 374 (2017)








\end{thebibliography}

\end{document}